# Correlations and dynamics of spins in an XY-like spin-glass $(Ni_{0.4}Mn_{0.6})TiO_3$ single crystal system

R. S. Solanki,[1,§] S. -H. Hsieh,[1] C. H. Du,[1] G. Deng,[2] C. W. Wang,[2,3] J. S. Gardner,[3] H. Tonomoto,[4] T. Kimura,[4] W. F. Pong[1,*]

[1] Department of Physics, Tamkang University, Tamsui 251, Taiwan

[2] Australian Nuclear Science and Technology Organization, Lucas Height, NSW 2233, Australia

[3] Neutron Group, National Synchrotron Radiation Research Center, Hsinchu 30076, Taiwan

[4] Division of Materials Physics, Graduate School of Engineering Science, Osaka University, Toyonaka, Osaka 560-8531, Japan

## Abstract

Elastic and inelastic neutron scattering (ENS and INS) experiments were performed on a single crystal of $(Ni_{0.4}Mn_{0.6})TiO_3$ (NMTO) to study the spatial correlations and dynamics of spins in the XY-like spin-glass (SG) state. Magnetization measurements reveal signatures of SG behavior in NMTO with a freezing temperature of $T_{SG} \sim 9.1$ K. The ENS experiments indicated that the intensity of magnetic diffuse scattering starts to increase around 12 K, which is close to $T_{SG}$. Also, spin-spin correlation lengths ($\xi$) at 1.5 K are approximately *(21±1) and (73±2)* Å in the interlayer and the in-plane directions, respectively, demonstrating that magnetic correlations in NMTO exhibit quasi two-dimensional antiferromagnetic order. In addition, critical exponent ($\beta$) is determined to be 0.37±0.02 from the intensity of magnetic diffuse scattering confirms the XY-like SG state of NMTO. INS results show quasi-elastic neutron scattering (QENS) profiles below $T_{SG}$. The life-time of dynamic correlations, $\tau \sim \hbar/\Gamma_L$, obtained from the half width at half maximum of the Lorentzian ($\Gamma_L$) QENS profiles, are approximately *(16±1) and (16±2) ps* at 10 K for two positions (0.00, 0.00, 1.52) and (0.01, 0.01, 1.50), respectively. Therefore, our experimental findings demonstrate that short-range-ordered antiferromagnetic clusters with short-lived spin correlations are present in the XY-like SG state of NMTO.

## I. Introduction

Spin-glass (SG) systems are examples of frustrated magnetism and have attracted much interest as they exhibit a wide range of interesting physical phenomena, such as history-dependence and divergent nonlinear magnetic susceptibilities.[1-3] Typically, the combination of competing exchange interactions and either site or bond disorder leads to an SG state. However, in some stoichiometric intermetallic compounds, such as $PrAu_2Si_2$[4] and $PrRuSi_3$[5], which exhibit neither static disorder nor a geometrically frustrated lattice, transition to the SG state has been observed. Inelastic neutron scattering (INS) studies have shown that the SG behavior in these systems arises from dynamic fluctuations of the crystal field levels.[4] These fluctuations destabilize the induced magnetic moments and frustrate the development of long-range magnetic ordering. To date, neutron scattering has proven to be a powerful tool for elucidating the nature of static and dynamic magnetic correlations in the SG state of several disordered alloys and in geometrically frustrated compounds such as $Cu_xMn_{1-x}$,[6] $Fe_xAl_{1-x}$,[7] $Fe(Ni_{1-x}Mn_x)$,[8,9] $Y_2Mo_2O_7$[10] and $La_2(Cu, Li)O_4$[11]. However, the different behaviors depending on whether one is dealing with the Ising-, XY- or Heisenberg-type of SG systems is still a subject of research.[12] $(Fe_{0.50}, Mn_{0.50})TiO_3$ is reported to behave like a typical Ising SG system[13,14] while the dilute magnetic alloys such as $Cu_xMn_{1-x}$, $Ag_xMn_{1-x}$ or $Au_xFe_{1-x}$ [15,16] are known as canonical Heisenberg-like SG systems [17]. However, true XY spin glass systems are rare. The chiral-glass superconductors are close to the XY SG systems but there SG state is often referred to as orbital glass state[17,18] because it arises due to orbital moments rather than spins. Also, Mathieu et al. report that the SG state in $Pr_{1-x}Ca_{1+x}MnO_4$ arises due to the presence of passive $e_g$ orbitals.[19]

In the present work, we focus on an XY-like SG system $(Ni_{0.4}Mn_{0.6})TiO_3$ (NMTO) that exhibits memory and relaxation effects and has more recently been reported to exhibit linear magnetoelectric (ME) coupling.[20-22] $(Ni_xMn_{1-x})TiO_3$ with $0.40 \leq x \leq 0.48$ has an ilmenite crystal structure with a rhombohedral $R\bar{3}$ space group (centrosymmetric). In the crystal structure of $(Ni_xMn_{1-x})TiO_3$, a magnetic (Ni, Mn) plane and a non-magnetic Ti plane are alternate along the hexagonal *c*-axis, separated by oxygen layers. Ni and Mn ions in each (Ni, Mn) plane form a honeycomb lattice.[20,21] The parent compounds $NiTiO_3$ and $MnTiO_3$ are antiferromagnetic and have the ilmenite crystal structure (space group $R\bar{3}$). Figures 1(a) and (b) depict the schematic magnetic structures of $NiTiO_3$ and $MnTiO_3$ after Yamaguchi *et al.*[20] In $NiTiO_3$, the spin arrangement is ferromagnetic within the hexagonal *c*-layer, but adjacent layers are coupled antiferromagnetically with Néel temperature, $T_N \approx 23$ K[20,22] and the spin easy axis is perpendicular to the *c*-axis. The magnetic structure of $NiTiO_3$ is *A*-type with magnetic wave vector $q = (0, 0, 1.5)$.[22,23] However, in $MnTiO_3$, spin ordering is antiferromagnetic in both inter- and intra-layer directions with $T_N \approx 64$ K and the spin easy axis is parallel to the *c*-axis. $MnTiO_3$ has a *G*-type magnetic structure with magnetic wave vector $q = (0, 0, 0)$.[22,23] $MnTiO_3$ reportedly exhibits ME coupling owing to its magnetic symmetry[24] and a spin-flop transition[25]. The different spin arrangements and easy axes of the parent compounds compete in the mixed compound so frustration of the exchange interactions in NMTO gives rise to its SG behavior at low temperatures. Competitions between spin anisotropies result in effective XY-easy plane anisotropy.[26,27]

In this work, x-ray diffraction, magnetization, neutron powder diffraction, elastic and inelastic neutron scattering (ENS and INS) experiments were performed on single crystal and powder samples of NMTO to elucidate the correlations and dynamics of spins in the SG state. Single crystal x-ray diffraction verifies the high quality of the sample. Rietveld refinement of the powder x-ray diffraction pattern reveals that the sample has a single phase. Temperature-dependent dc magnetization measurements provide evidence of a SG transition in the sample at $T_{SG}$~ 9.1 K with a Curie-Weiss temperature ($\theta_{CW}$) of -91.28 K. Neutron powder diffraction patterns show strong magnetic diffuse scattering at finite ***Q*** which is not coinciding with the nuclear Bragg peaks. These features reveal that short-range antiferromagnetic spin-spin correlations exist in the SG. ENS studies of the single crystal NMTO indicate that the magnetic correlations are quasi two-dimensional (2D). INS studies reveal that the life-times ($\tau$) of the dynamic correlations are similar to those of SG systems.[1-3]

## II. Experimental

Single crystals of NMTO were prepared by the floating zone method, as described elsewhere.[20,21] X-ray diffraction patterns were obtained using an in-house x-ray diffraction facility with Cu $K_\alpha$ radiation. Magnetic measurements were made on the polycrystalline sample of NMTO using a superconducting quantum interference device magnetometer. Neutron powder diffraction patterns were obtained at a wavelength of 4.22 Å using the WOMBAT diffractometer at ANSTO, Australia, with a pyrolytic graphite (PG) monochromator: a Be filter was used to remove higher-order contamination. ENS and INS experiments were performed on a single crystal of NMTO on a cold neutron triple-axis spectrometer-SIKA at ANSTO, Australia[28]. SIKA is equiped with a PG (002) monochromator and analyzer. For the current experiment, SIKA run in a constant $E_f$ mode, with 40'-40'-40'-40' for the pre-mono, pre-sample, pre-analyzer and pre-detector collimators, respectively. Some of the data were collected at a constant $E_f$= 5.5 meV with a PG filter. Other data were measured by using a cooled Be-filter at $E_f$= 3.7 meV. Both PG filter and Be filter were used in order to suppress the high order contamination. The temperature of the sample was controlled using a standard cryogenic He orange cryostat. Rietveld refinements of the x-ray powder diffraction data were carried out using the FULLPROF software suite.[29]

### III. Results and discussion

Rietveld analysis of the x-ray powder diffraction pattern that was obtained at room temperature was carried out with $R\bar{3}$ space group model and presented in Fig. 1(c). All the observed peaks in the diffraction pattern were accounted for using the single phase $R\bar{3}$ space group, except for those associated with a negligibly small amount of impurity in the form of the rutile phase of $TiO_2$ (JCPDS No.: 34-0180), at 2θ~27° marked as *. The refined lattice parameters of NMTO were $a= b= 5.125$ Å and $c= 14.120$ Å in hexagonal structure with agreement factors $\chi^2= 3.79$ and $R_{wp}= 7.40$. The upper inset in Fig. 2(a), presents the x-ray diffraction profiles for the (113) reflection of the single crystal of NMTO. The small FWHM~ $0.16^0$ verifies the high quality of the crystal. To determine the SG transition temperature, the temperature-dependence of the zero-field cooled (ZFC) and field-cooled (FC) *dc* magnetic susceptibility (χ) of a powdered sample of NMTO in a field of 0.5mT was plotted in the lower inset in Fig. 2(a). The irreversibility of ZFC and the FC χ below $T_{SG}$ ~ 9.1 K in the low field is the signature of the SG behavior of NMTO. Further, Fig. 2(b) shows the temperature-dependence of the real part of *ac* magnetic susceptibility (χ') for H//*Y-axis* (⊥ *c*-axis) and H//*Z-axis* (∥ *c*-axis). The position of maximum of χ' shifts to the high temperature with increasing frequency which is the characteristics of conventional spin glasses.[1-3] Similar, behavior has been previously reported by Yamaguchi *et al.*[20] for the SG state of NMTO. These features verify that the SG state of NMTO is stable below ~ 9.1 K. The high-temperature χ data that were collected in a magnetic field of 1 T are consistent with Curie-Weiss behavior, $\chi(T)= C/(T-\theta_{CW})$. The linear fit (solid line) of $\chi^{-1}$ as a function of T from 120 to 300 K [Fig. 2(a)] yields a Curie-Weiss temperature ($\theta_{CW}$) of -91.28 K and an effective moment ($\mu_{eff}$) of 4.73 $\mu_B$. The strongly negative value of $\theta_{CW}$ reveals predominant antiferromagnetic exchange interactions in NMTO.[5,10] The experimentally obtained value of $\mu_{eff}$ lies between the theoretically expected spin only magnetic moment, $g\sqrt{S(S+1)}\mu_B$ of 2.828 $\mu_B$ for $Ni^{2+}$ (S= 1) and 5.916 $\mu_B$ for $Mn^{2+}$ (S= 5/2) free ions. To analyze the short-range-ordered SG state of NMTO, neutron powder diffraction patterns of NMTO at 1.6 K and 20 K are displayed in Fig. 2(c). A very broad peak at finite $Q \approx$ 0.67 Å$^{-1}$ (indicated by an arrow) is observed in the neutron powder diffraction pattern that was obtained at 1.6 K, which is below SG transition temperature of $T_{SG} \sim$ 9.1 K. Strong diffuse scattering at finite $Q$ which in not coinciding with nuclear Bragg peak along with magnetization measurement indicates that short-range spin-spin correlations are antiferromagnetic in nature.[7-9,27] All of the other peaks in the pattern are indexed to a rhombohedral $R\bar{3}$ space group in the hexagonal phase. A peak of low intensity that is marked by * at $Q \approx$ 1.41 Å$^{-1}$ is associated with λ/2 contamination from the PG monochromator. To investigate the systematic temperature-dependence of the correlations and dynamics of spins in the SG state, ENS and INS measurements were performed on a single crystal sample of NMTO.

Figure 3(a) presents mesh scans of the single crystal of NMTO at $E_f$ = 5.5 meV around the (0, 0, 1.5) position at a temperature of 1.5 K.[27] The reciprocal lattice point (0, 0, 1.5) represents the first antiferromagnetic peak in the parent compound, $NiTiO_3$.[22,23] The temperature dependence of this scattering in NMTO indicates that it is magnetic in origin similar to that observed in the neutron powder diffraction pattern. To eliminate contributions from nonmagnetic nuclear scattering and higher-order contaminations, data obtained at 30 K [Fig. 3(b)] were subtracted from those obtained at 1.5 K [Fig. 3(a)]. Figure 3(c) displays the pattern following subtraction. The anisotropy of the patterns in the [0, 0, L] and [H, H, 1.5] directions is the signature of two spin-spin correlation lengths, $\xi$. Although some elastic neutron scattering results around (0, 0, 1.5) reciprocal lattice point has been reported by Kawano *et al.*[27] previously, such anisotropy has not before been observed in the SG state of NMTO. To determine reliably spatial correlation, $\xi$ value in both directions and their dependence on temperature around $T_{SG}$, ENS data were obtained systematically from 1.5 to 30 K.

Figure 4 displays the temperature-dependent ENS profiles for NMTO around the (0, 0, 1.5) reciprocal lattice point in (a) the inter-plane/layer ([0, 0, L]) direction and (b) the in-plane ([H, H, 1.5]) direction at several temperatures around $T_{SG}$, obtained at $E_f = 3.7$ meV. The scattering pattern includes a central Bragg-like peak and background (*BG*). As the temperature increases from 1.5 K to above $T_{SG}$, the intensity of central peak decreases, reaching close to zero at temperatures of above 12 K. The observed patterns are well fitted by a Lorentzian function, which represents magnetic diffuse scattering from the sample with the addition of a constant *BG* term. Therefore, the total scattering function is:

$$I(Q) = BG + \frac{A_L}{\pi} \frac{\kappa_L}{[\kappa_L^2 + Q^2]}$$

where $A_L$ and $\kappa_L$ represent the integrated intensity and HWHM of the Lorentzian function,[11,22] respectively. Figures 4(a) and (b) show fitting results using the above function. Different value of $\xi$ ($\xi \approx \kappa_L^{-1}$) were obtained in the two directions. At 1.5 K, in the inter-plane/layer direction, the $\xi$ is approximately *(21±1)* Å, while in the in-plane direction it is approximately *(73±2)* Å. Figure 4(a) reveals that the inter-plane/layer $\xi$ exceeds the distance between the neighboring Mn/Ni layers ($c$/3= 4.7066Å). Generally, a small ξ corresponds to the nanometer-scale spin clusters.[7-9] These features further indicate that the magnetic spin-spin correlations are quasi 2D.[11] The insets in Figs. 4(a) and (b) indicate the strong temperature-dependence of ξ and $A_L$. Increasing the temperature reduces drastically both ξ and $A_L$, which are close to zero at 20 K. The increase in $A_L$ below $T_{SG}$ suggests that the number of antiferromagnetically correlated clusters increases as the temperature decreases.[9] Further, integrated intensity of the magnetic diffuse scattering is proportional to the square of local sublattice magnetization (order parameter).[1-3] Near the SG transition it follows a power-law, $I \propto (T_{SG}-T)^{2\beta}$ where β is the critical exponent related to the order parameter. The integrated intensities along two directions in the temperature range from 1.5 to 12 K have been fitted with the power-law using β and $T_{SG}$ as fitting parameters. The continuous lines in the insets correspond to the fitting with β≈ (0.37±0.02) and $T_{SG}$≈ (12.4±0.1) K. The value of exponent, β is close to XY model[1-3,30] and therefore confirms the XY-like nature of SG state of NMTO. The value obtained for $T_{SG}$ is higher than that obtained from the magnetization measurement due to the non-zero integrated intensity even above 9.1 K and can be assigned to the slow dynamics of SG systems. To establish the dynamics of the experimentally observed short-range spin-spin correlations, INS data were collected slightly away from the (0, 0, 1.5) point to avoid the strong elastic scattering but sufficiently close to obtain a reasonable signal from the short-range spin-spin correlations.[31]

Figures 5(a) and (b) present the INS spectra as functions of energy transfer ($E$) for NMTO from 1.5 to 50 K; two positions (a) (0, 0, 1.52) and (b) (0.01, 0.01, 1.50) corresponding to transverse and longitudinal displacements from (0, 0, 1.50) reciprocal lattice point are measured. Owing to the very low intensity of quasi-elastic neutron scattering (QENS), the insets in Figs.

5(a) and (b) display a magnified view of the tail region. QENS profiles are indicated by arrows and can be modeled using a Lorentzian function. The total scattering function is:

$$I(\omega) = BG + Y_G + \frac{A_L}{\pi} \frac{\Gamma_L}{[\Gamma_L^2 + (\hbar\omega)^2]}$$

where $BG$ is the background, $Y_G$ represents the resolution-limited elastic Gaussian component and $A_L$ & $\Gamma_L$ are the integrated intensity and HWHM of the Lorentzian function,[7-9, 32] respectively. Figures 6(a) and (b) present the results of fitting at 1.5 K using the above function for the two positions. INS data at 50 K are closely described using a single Gaussian function, but the tails that arise from QENS are well fitted using a single Lorentzian component. Therefore, all INS data were analyzed using the aforementioned function. The HWHMs of the elastic Gaussian components (instrumental resolution) are (0.029±0.007) and (0.030±0.007) meV for (0, 0, 1.52) and (0.01, 0.01, 1.50) positions. These HWHMs are determined using vanadium scans under similar conditions and held fixed during fitting. Figures 7(a) and (b) plot the variations of the obtained parameters $A_L$ and $\Gamma_L$ with temperature for the two positions. Figures 7(c) and (d) show the elastic Gaussian contributions ($A_G$ and $\Gamma_G$). The $A_G$ of the elastic Gaussian component remains nearly constant for T≥ 20 K but starts to increase rapidly as the temperature falls below 12 K. These figures indicate that for the two positions, the integrated intensity of the quasi-elastic Lorentzian component is zero for T≥ 20 K, but that of the elastic Gaussian component is almost constant for T≥ 20K with a HWHM that is still comparable to the instrumental resolution. The temperature-dependence of the integrated intensity of the QENS ($A_L$) profiles is similar to those observed for some other SG systems and attributed to the slow dynamics of the spin correlations.[7,11,33] However, the spin-relaxation rates ($\Gamma_L$) decrease drastically as the temperature falls below $T_{SG}$. The rapid decrease of the relaxation rate below $T_{SG}$ has been observed in many

SG systems[10,34,35] and saturates for T< 5 K to a value that is close to the resolution limit of the instrument. The spin-relaxation rates at 10 K for (0, 0, 1.52) and (0.01, 0.01, 1.50) positions are $\Gamma_L$ ~ (0.21±0.01) and ~(0.16±0.02) meV, respectively. The life-time of the dynamic correlations, $\tau \sim \hbar/\Gamma_L$, are approximately *(16±1) and (16±2)* ps at 10 K for the two positions, which are comparable to those of typical SG systems.[1-3,35] However, the abrupt increase in the intensity of the elastic Gaussian component as the temperature falls below 12 K reveals that there are two magnetic contributions in the NMTO: first, short-range spin correlations give rise to QENS and can be described using a Lorentzian function and second, a slower component that appears static within our instrumental resolution can be described using the Gaussian function.[6,10,11]

As mentioned above, the magnetic structure of $NiTiO_3$ is *A*-type and its spin arrangement is ferromagnetic in the hexagonal *c*-layer, but adjacent layers are coupled antiferromagnetically. In contrast, the magnetic structure of $MnTiO_3$ is *G*-type, and the spin ordering is antiferromagnetic in both inter- and intra-layer directions. The different spin arrangements of the parent compounds compete in the mixed compound so frustration of the exchange interactions in NMTO gives rise to SG behavior at low temperatures. Meanwhile, as discussed in the introduction, the SG state can arise from a combination of frustration of long-range magnetic interactions and chemical disorder.[1-3] However, we conceive that the XY-like SG state of NMTO is also associated with the electronic and orbital properties of $Ni^{2+}$ and $Mn^{2+}$ ions particular the former one at $T_{SG}$, owing to the strong spin-orbital coupling effects as observed in $Eu_{0.5}Sr_{1.5}MnO_4$ [17] and $Pr_{1-x}Ca_{1+x}MnO_4$ [19]. Accordingly, the electronic and orbital properties at the $Ni^{2+}$ and $Mn^{2+}$ sites of NMTO at/near $T_{SG}$ must be investigated to elucidate the critical role of bond/site disorder and orbital properties in the stabilization of the SG state in the NMTO. This issue is currently being studied using x-ray absorption/resonant inelastic x-ray scattering/soft x-ray scattering and will be reported elsewhere.

## IV. Conclusions

In summary, a XY-like SG system, NMTO, with ME coupling was studied using various experimental techniques. Magnetization measurements show that the SG state of NMTO is stable below $T_{SG} \sim$ 9.1 K. Neutron powder diffraction experiments verify strong magnetic diffuse scattering around the (0, 0, 1.5) reciprocal lattice point at 1.6 K which correlates with the AFM zone centre of the parent compound. ENS experiments on a single crystal of NMTO also reveal magnetic diffuse scattering around the (0, 0, 1.5) reciprocal lattice point for T≤ 12K. The small values of ξ provide the evidence that magnetic correlations are quasi 2D. Moreover, critical exponent (β) obtained from the intensity of magnetic diffuse scattering lies close to the XY spin-glass system. INS results suggest that the dynamics of the spins start to freeze for the both positions below $T_{SG}$, saturating at values that are close to the instrumental resolution. The life-time of the dynamic correlations, $\tau \sim \hbar/\Gamma_L$ are approximately *(16±1) and (16±2)* ps at 10 K for the (0, 0, 1.52) and (0.01, 0.01, 1.50) positions. Therefore, the detailed investigation of temperature-dependent magnetization, powder neutron diffraction, ENS and INS data herein reveal that short-range-ordered antiferromagnetic clusters with slow spin dynamics are characteristics of the SG state of NMTO.

*Author to whom all correspondence should be addressed; Electronic mail: wfpong@mail.tku.edu.tw

§Present affiliation: Centre of Material Sciences, Institute of Interdisciplinary Studies, University of Allahabad, Allahabad-211002, Uttar Pradesh, India

**Acknowledgement:**

The author (W.F.P.) would like to thank the Ministry of Science and Technology (MoST) of the Taiwan for financially supporting this research under Contract Nos. NSC 102-2112-M032-007-MY3 and NSC 102-2632-M032-001-MY3. W.F.P., R.S.S. and S.H.H. are thankful to the NSRRC, Taiwan for providing financial assistance to visit SIKA, ANSTO, Australia to collect the neutron scattering data under proposal nos. N-2014-3-005 and N-2015-1-016. W.F.P. also acknowledges the support of SIKA-Neutron Program of MoST, Taiwan and staffs of ANSTO, Australia. H.T. and T.K. were in part supported by JSPS KAKENHI Grant No. 24244058. R.S.S. is thankful to Department of Science and Technology (DST), India for DST INSPIRE Faculty award (DST/INSPIRE/04/2015/002300).

**References:**

1. K. Binder and A. P. Young, Rev. Mod. Phys. **58,** 801 (1986).
2. K. H. Fischer and J. A. Hertz, *Spin Glasses* (Cambridge Univ. Press, Cambridge, 1991).
3. J. A. Mydosh, *Spin glasses* (Taylor and Francis, London, 1993).
4. E. A. Goremychkin, R. Osborn, B. D. Rainford, R. T. Macaluso, D. T. Adroja & M. Koza, Nature Physics **4,** 766 (2008).
5. V. K. Anand, D. T. Adroja, A. D. Hillier, J. Taylor, and G. André, Phys. Rev. B **84**, 064440 (2011).
6. A. P. Murani and A. Heidemann, Phys. Rev. Lett. **41,** 1402 (1978).
7. W. Bao, S. Raymond, S. M. Shapiro, K. Motoya, B. Fåk, and R. W. Erwin, Phys. Rev. Lett. **82,** 4711 (1999).
8. K. Motoya, S. Kubota and S. M. Shapiro, J. Mag. and Mag. Mat. **140–144,** 75-76 (1995).
9. K. Motoya, Y. Muro and T. Igarashi, J. Phys. Soc. Jpn. **78 (5),** 054711 (2009).
10. J. S. Gardner, B. D. Gaulin, S.-H. Lee, C. Broholm, N. P. Raju, and J. E. Greedan, Phys. Rev. Lett. **83**, 211 (1999).

11. W. Bao, Y. Chen, Y. Qiu, and J. L. Sarrao, Phys. Rev. Lett. **91,** 127005 (2003); Y. Chen, W. Bao, Y. Qiu, J. E. Lorenzo, J. L. Sarrao, D. L. Ho, and Min Y. Lin, Phys. Rev. B **72**, 184401 (2005).

12. H. Kawamura, J. Phys.: Condens. Matter 23, 164210 (2011); D. L. Stein and C. M. Newman, *Spin Glasses and Complexity* (Princeton University Press, 2013).

13. D. Akahoshi, M. Uchida, Y. Tomioka, T. Arima, Y. Matsui, and Y. Tokura, Phys. Rev. Lett. **90,** 177203 (2003).

14. E. Torikai, A. Ito, I. Watanabe, and K. Nagamine, Physica B **374–375,** 95–98 (2006).

15. L. P. Lévy, Phys. Rev. B **38,** 4963 (1988).

16. D. Petit, L. Fruchter, and I. A. Campbell, Phys. Rev. Lett. **88,** 207206 (2002).

17. R. Mathieu, A. Asamitsu, Y. Kaneko, J. P. He, and Y. Tokura, Phys. Rev. B **72,** 014436 (2005).

18. H. Kawamura, J. Phys. Soc. Jpn. **64,** 711 (1995).

*19.* R. Mathieu, J. P. He, X. Z. Yu, Y. Kaneko, M. Uchida, Y. S. Lee, T. Arima, A. Asamitsu and Y. Tokura, EPL **80,** 37001 (2007).

20. Y. Yamaguchi, T. Nakano, Y. Nozue, and T. Kimura, Phys. Rev. Lett. **108,** 057203 (2012).

21. Y. Yamaguchi and T. Kimura, Nat. Comm. **4,** 2063 (2013).

22. S. Chi, F. Ye, H. D. Zhou, E. S. Choi, J. Hwang, H. Cao, and J. A. Fernandez-Baca, Phys. Rev. B **90,** 144429 (2014).

23. G. Shirane, S. J. Pickart, and Y. Ishikawa, J. Phys. Soc. Jpn. **14,** 1352 (1959).

24. N. Mufti, G. R. Blake, M. Mostovoy, S. Riyadi, A. A. Nugroho, and T. T. M. Palstra, Phys. Rev. B **83,** 104416 (2011).

25. H. Yamauchi, H Hiroyoshi, M. Yamada, H. Watanabe and H. Takei, J. Mag. and Mag. Mat. **31**, 1071 (1983).

26. H. Yoshizawa, H. Kawano, H. Mori, S. Mitsuda and A. Ito, Physica B **180-181**, 94-96 (1992); A. Ito, H. Kawano, H. Yoshizawa and K. Motoya, J. Mag. & Mag. Mat. **1637-1638,** 104-1117 (1992).

27. H. Kawano, H. Yoshizawa, A. Ito, and K. Motoya, J. Phys. Soc. Japan **62,** 2575 (1993).

28. C. M. Wu, G. Deng, J. S. Gardner, P. Vorderwisch, W. H. Li, S. Yano, J. C. Peng, and E. Imamovic, JINST **11,** P10009 (2016).

29. J. Rodriguez-Carvajal FULLPROF, *a Rietveld refinement and pattern matching analysis program,* Laboratory Leon Brillouin (CEA-CNRS); 2011. Available at: https://www.ill.eu/sites/fullprof/.

30. T. E. Mason, B. D. Gualin and M. F. Collins, Phys. Rev. B **39**, 586 (1989); R. Mathieu, A. Asamitsu, Y. Kaneko, J. P. He and Y. Tokura, Phys. Rev. B **72**, 014436 (2005).

31. S. Chi, F. Ye, P. Dai, J. A. Fernandez-Baca, Q. Huang, J. W. Lynn, E. W. Plummer, R. Mathieu, Y. Kaneko, and Y. Tokura, Proc. Nat. Acad. Sci. **104**, 10796 (2007).

32. J. Qvist, H. Schober, and B. Halle, J. Chem. Phys. **134**, 144508 (2011).

33. B. J. Sternlieb, G. M. Luke, Y. J. Uemura, T. M. Riseman, J. H. Brewer, P. M. Gehring, K. Yamada, Y. Hidaka, T. Murakami, T. R. Thurston, and R. J. Birgeneau, Phys. Rev. B **41,** 8866 (1990).

34. S.-H. Lee, C. Broholm1, G. Aeppli, A. P. Ramirez, T. G. Perring, C. J. Carlile, M. Adams, T. J. L. Jones and B. Hessen, Europhys. Lett., **35** (2), pp. 127-132 (1996).

35. X. Lu, D. W. Tam, C. Zhang, H. Luo, M. Wang, R. Zhang, L. W. Harriger, T. Keller, B. Keimer, L.-P. Regnault, T. A. Maier, and P. Dai, Phys. Rev. B **90,** 024509 (2014).

**Figure captions**

**Fig. 1:** Magnetic structures of (**a**) NiTiO$_3$ (*A*-type) with magnetic wave vector, *q*= (0, 0, 1.5) and (**b**) MnTiO$_3$ (*G*-type) with magnetic wave vector, *q*= (0, 0, 0) after Yamaguchi *et al*. [Ref. 13]. (**c**) Observed (red dots), calculated (black continuous line) and difference (blue continuous line) patterns obtained following Rietveld refinement of room-temperature x-ray power diffraction pattern of NMTO. Vertical bars indicate Bragg positions. Small peak at 2θ~ 27$^0$ indicated by * represents impurity in form of rutile phase of TiO$_2$.

**Fig. 2:** (**a**) Temperature-dependence of the inverse magnetic susceptibility of polycrystalline NMTO sample in 10 kOe magnetic field under field-cooled conditions. Red continuous line represents Curie-Weiss fit from 120-300K. Upper inset: x-ray diffraction profile of (113) reflection of single crystal of NMTO, Lower inset: temperature-dependence of ZFC and FC magnetization of polycrystalline sample in 5.0 Oe magnetic field, (**b**) Variation of real component χ' of ac magnetic susceptibility for *H // Y*-axis and *H // Z*-axis (*H*=5.0 Oe) at 10, 100 and 1000 Hz frequencies with temperature for single crystal of NMTO and (**c**) Neutron powder

diffraction patterns obtained at temperatures of 1.6 K and 20 K. Arrow in figure indicates intense diffuse scattering at $Q \approx 0.67$ Å$^{-1}$ at 1.5 K. Small peak at $Q \approx 1.41$ Å$^{-1}$ marked with * arise from λ/2 contamination from PG monochromator.

**Fig. 3:** Mesh scans around (0, 0, 1.5) reciprocal lattice point at **(a)** 1.5 K and **(b)** 30 K at $E_f = 5.5$ meV. To remove nonmagnetic contributions, 30 K data were subtracted from 1.5 K data and results are shown in **(c)**.

**Fig. 4:** Temperature-dependence of ENS spectra in **(a)** the inter-layer/plane direction ([0, 0, L]) and **(b)** in-plane direction ([H, H, 1.5]) between 1.5 and 30 K. Continuous lines through dots represent Lorentzian curve fitting, discussed in text. Insets show correlation lengths and integrated intensities in the two directions. Dotted lines through data points are guide to eyes while continuous lines through the integrated intensities corresponds to the power-law fit, $I \sim (T_{SG} - T)^{2\beta}$ discussed in the text. Error bars size in insets are comparable to that of data points.

**Fig. 5:** INS spectra as a function of energy transfer ($E$) for NMTO from 1.5 to 50 K in **(a)** (0, 0, 1.52) and **(b)** (0.01, 0.01, 1.50) positions. Insets in Figs., **(a)** and **(b)** show magnified view that show QENS at temperatures from 1.5 to 12 K, indicated by arrows.

**Fig. 6:** Results of fitting of INS spectra for **(a)** (0, 0, 1.52) and **(b)** (0.01, 0.01, 1.50) positions at 1.5 K, using a combination of Gaussian and Lorentzian functions (continuous green line) with constant background; continuous pink line represents Gaussian component at 1.5 K. At 50 K, profile is Gaussian so a single Gaussian function (shaded blue region) is fitted to data, owing to incoherent scattering from the sample.

**Fig. 7:** Temperature-dependence of half width at half maximum ($\Gamma_L$ and $\Gamma_G$) and integrated intensities of Lorentzian (QENS) and Gaussian (central elastic) components in *(a, c)* (0, 0, 1.52) and *(b, d)* (0.01, 0.01, 1.50) positions. Integrated intensity of Lorentzian components due to QENS is zero at 20, 30 and 50 K. Error bars size for integrated intensity are comparable to data points.

**Fig. 1**

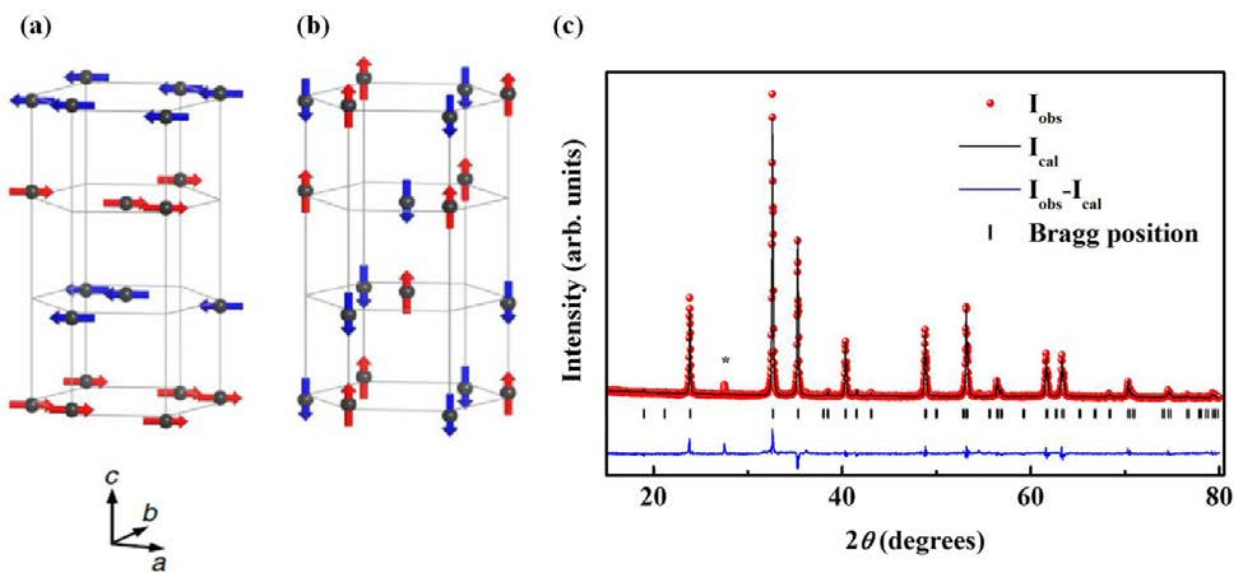

**Fig. 2**

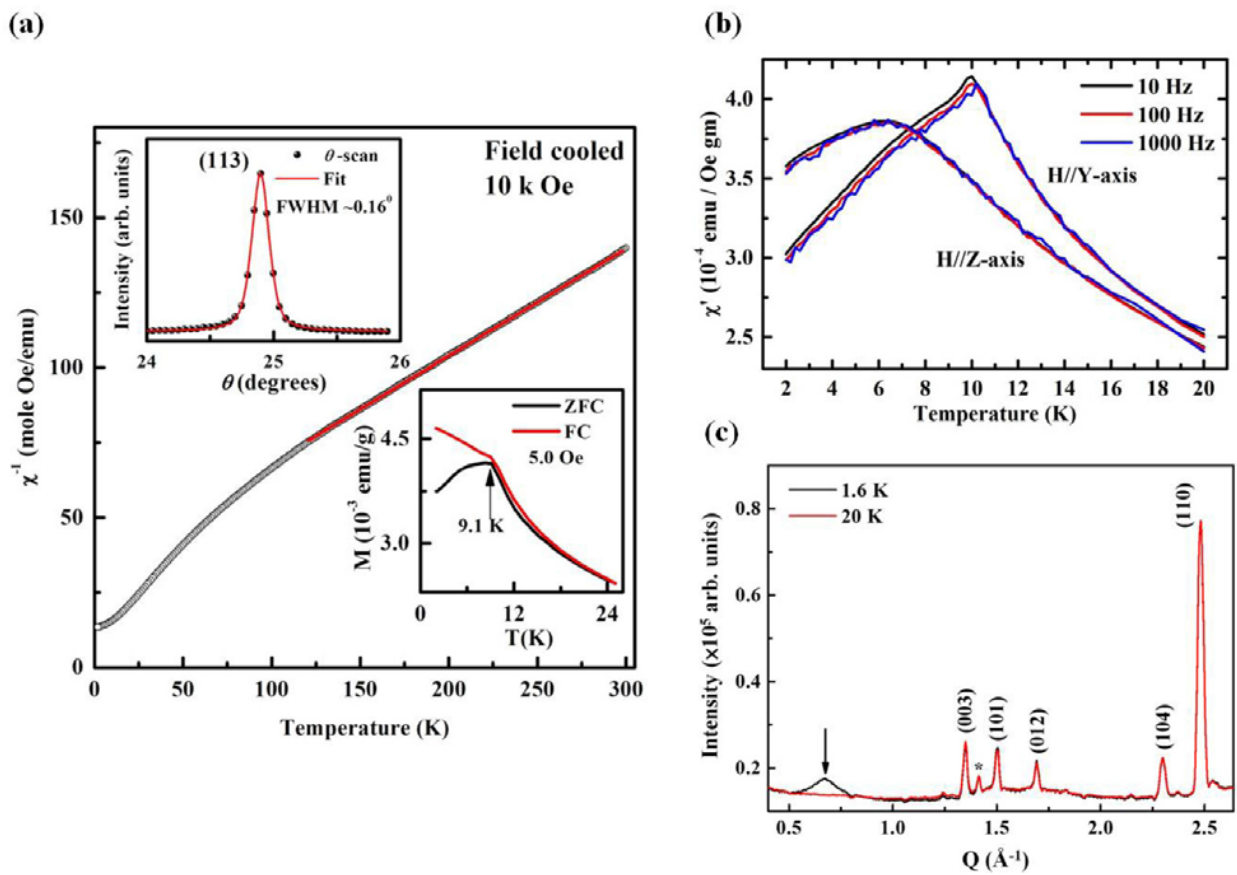

**Fig. 3**

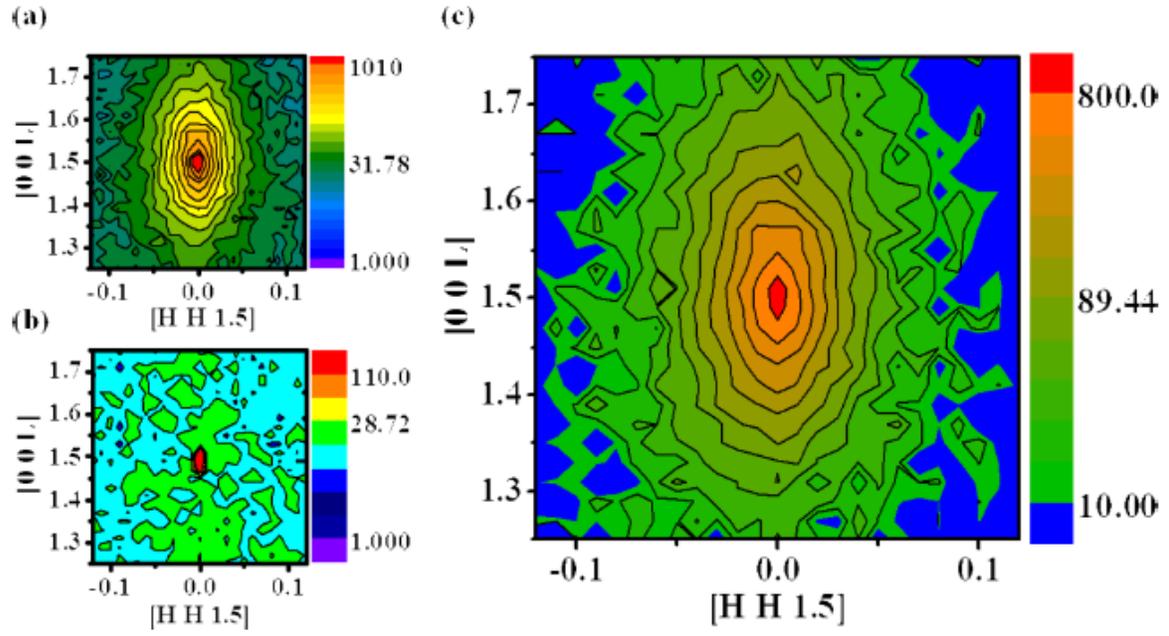

**Fig. 4**

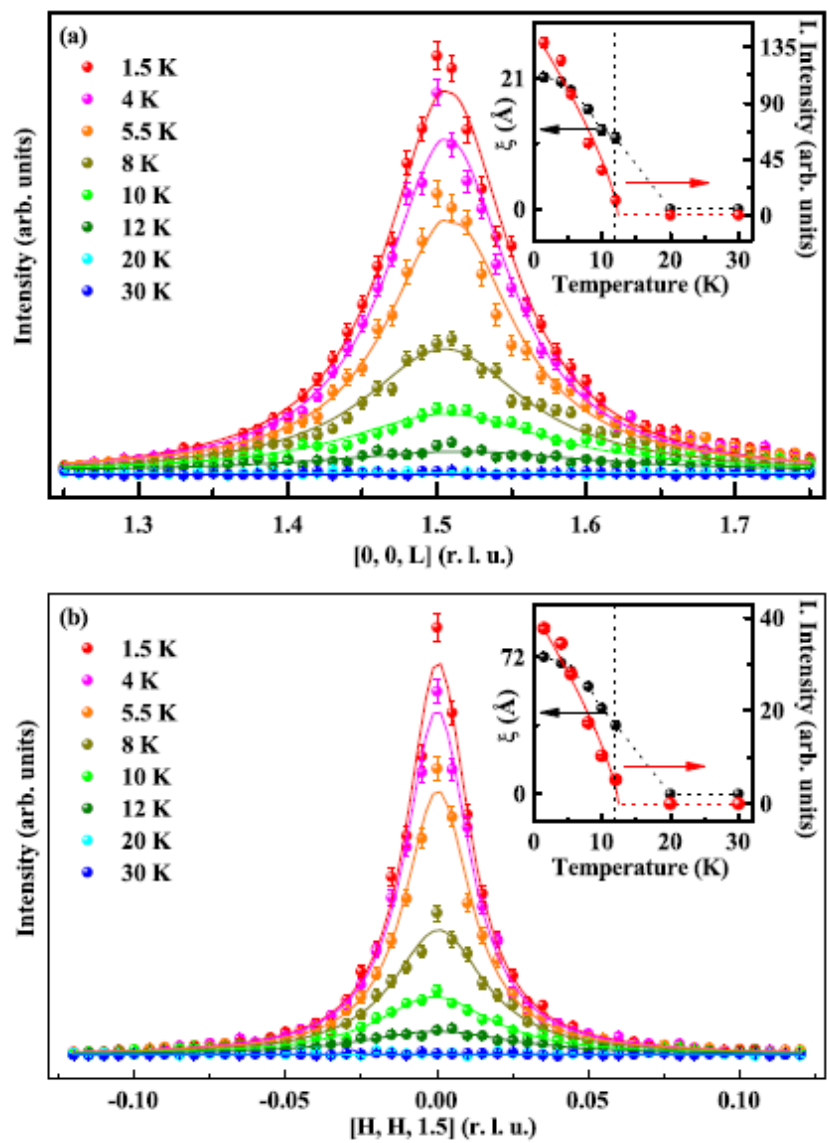

**Fig. 5**

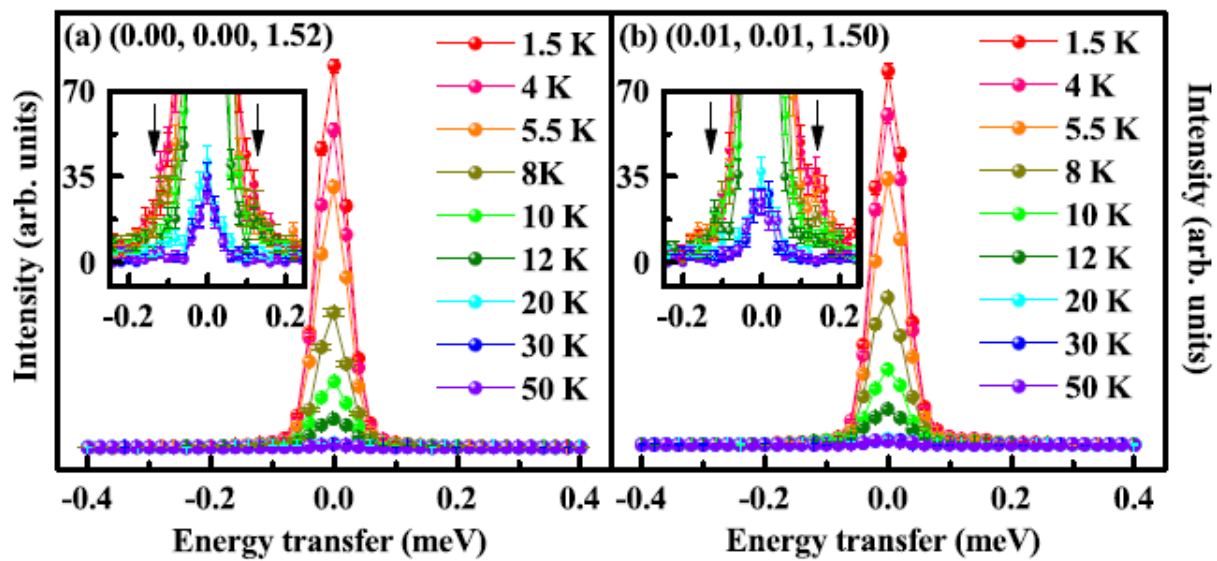

**Fig. 6**

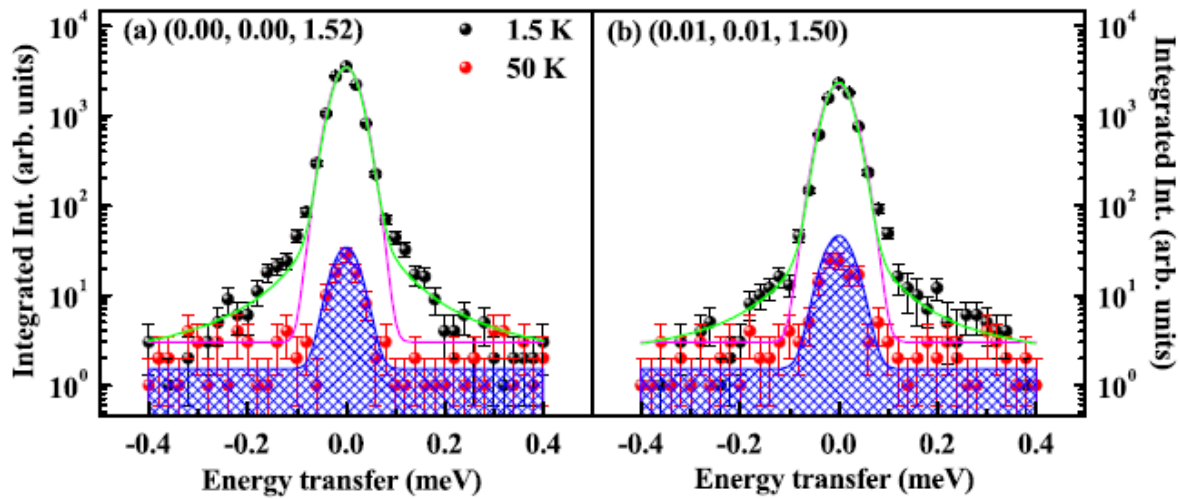

**Fig. 7**

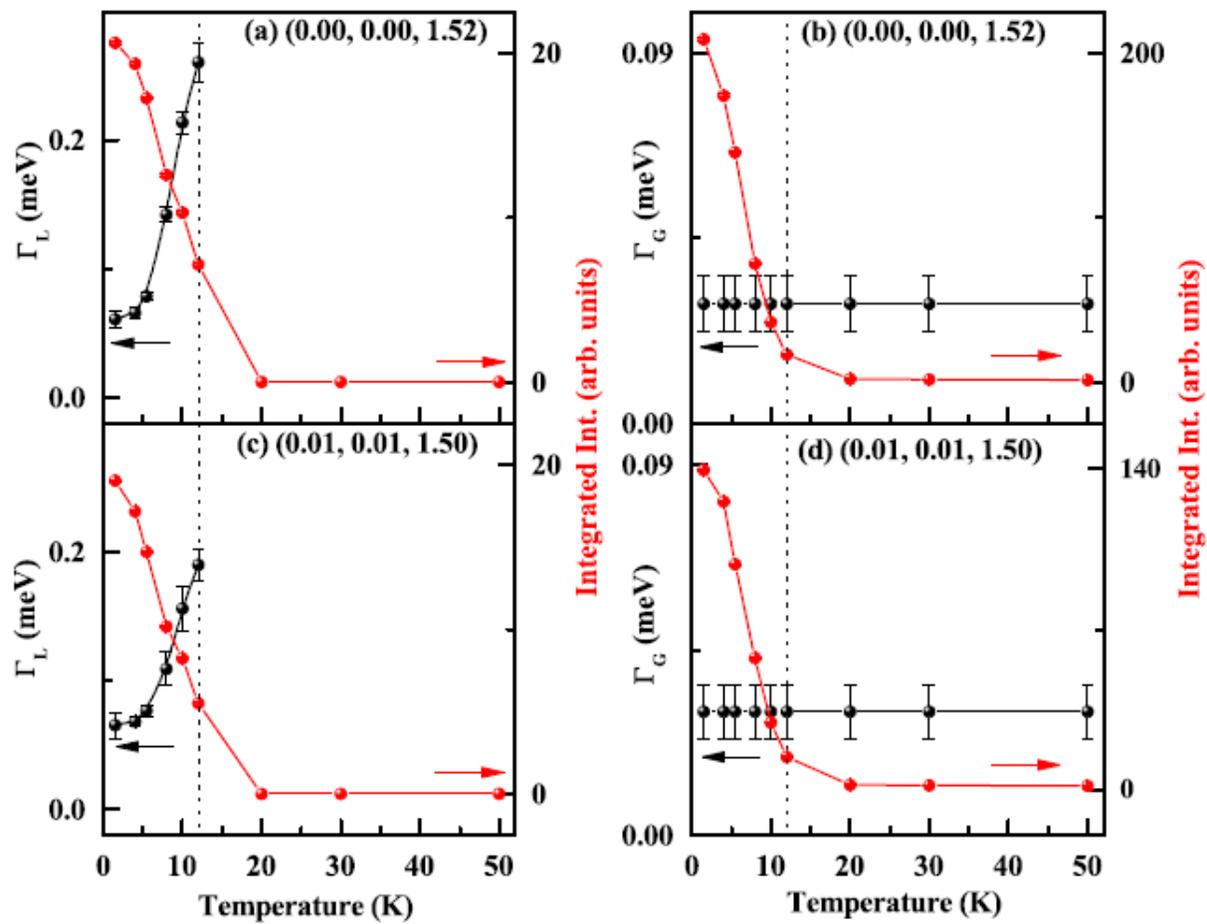